\documentclass[aps,prb,twocolumn,showpacs,showkeys]{revtex4-2}
\usepackage{graphicx}
\usepackage{dcolumn}
\usepackage{bm}
\usepackage{hyperref}
\usepackage{amsmath}
\usepackage{amssymb}
\usepackage{booktabs}
\usepackage{multirow}
\usepackage{diagbox}
\usepackage{makecell}
\usepackage{tabularx}
\usepackage{graphicx}
\usepackage{multirow}
\usepackage{longtable}
\usepackage{colortbl}
\usepackage{xcolor}
\usepackage{bbding}
\newcommand{\tabincell}[2]{\begin{tabular}{@{}#1@{}}#2\end{tabular}}
\usepackage[switch]{lineno} 
\hypersetup{colorlinks=true, citecolor=blue, urlcolor=blue, linkcolor=blue}
\usepackage{blindtext}

\makeatletter

\makeatletter

\begin{document}
	\title{A Universal Scaling Law for $T_c$ in Unconventional Superconductors}
	\author{Way Wang$^{1}$}
	\email{Corresponding author: waywang@zju.edu.cn}
	\author{Zhongshui Ma$^{1,2}$}
	\email{Corresponding author: mazs@pku.edu.cn}
	\author{Hai-qing Lin$^{1}$}
	\email{Corresponding author: hqlin@zju.edu.cn}
	\affiliation{$^{1}$ Institute for Advanced Study in Physics and School of Physics, Zhejiang University, Hangzhou 311058, China}
	\affiliation{$^{2}$ School of Physics, Peking University, Beijing 100871, China}
	
	\begin{abstract}
		
		Understanding the pairing mechanism of unconventional superconductors remains a core challenge in condensed matter physics, particularly the ongoing debate over whether the related effects caused by electron-electron interactions unify various unconventional superconductors (UcSs). To address this challenge, it is necessary to establish a universal quantitative relationship for the superconducting transition temperature ($T_c$), which can be directly obtained from experiments and correlated with microscopic parameters of different material systems. In this work, we establish a relation: $N_{\text{CP}}\cdot k_{B}T_{c}^\star = \alpha\cdot U $, where $\alpha = 1/(16\pi)$ is a universal constant, $k_B$ is the Boltzmann constant, $T_{c}^\star$ is the maximal $T_{c}$, $U$ is the on-site Coulomb interaction, and $N_{\text{CP}}$($\propto(\xi_0/a)^D$) quantifies the spatial extent of Cooper pairs ($\xi_0$) relative to lattice parameter ($a$) in $D$ dimensions. The validity of this scaling relationship is empirically demonstrated, across a four order-of-magnitude $T_c^\star$ range (0.08--133 K), by database from 173 different compounds spanning 13 different UcS families in over 500 experiments. The fact that the unified relationship is satisfied by different materials of different UcS families reveals that they may share superconducting mechanisms. In addition, the scaling relationship indicates the existence of a maximum $T_{c}^\star$ determined by the minimum $N_{\text{CP}}$, providing a benchmark for theoretical and experimental exploration of high-temperature superconductivity.
		
	\end{abstract}


\maketitle

\section{Introduction}

     The 1986 discovery of high-temperature superconductivity in cuprates by Bednorz and Müller\cite{Bednorz1986} suggested that the critical temperature could exceed the McMillan-Cohen predicted lattice-stability limits\cite{McMillan1968,Cohen1972}, and thereby challenge the phonon-mediated BCS pairing paradigm\cite{Bardeen1957}. This breakthrough highlighted the profound complexity of correlated electron systems\cite{Keimer2015} and drove decades of research into unconventional superconductors (UcSs, i.e., materials distinguished by their non-BCS character) that encompass cuprates, heavy-fermion compounds\cite{Steglich1979}, iron-based pnictides\cite{Kamihara2008}, twisted van der Waals heterostructures\cite{Cao2018}, nickelates\cite{Hwang2019,Sun2023}, and other exotic quantum condensates\cite{Norman2011,Stewart2017}. Elucidating the superconducting mechanisms governing UcSs remains a paramount challenge, further compounded by persistent controversy over whether a unified framework underlies these diverse materials\cite{Stewart2017,Norman2011, Xiang2025}. Resolving this debate requires establishing a universal, quantitative relationship for $T_c$---one directly derived from experimental observables and capable of correlating essential microscopic parameters across different material classes. 
    
     Early theoretical and experimental studies focused on establishing empirical relationships between $T_c$ and macroscopic properties. Uemura's seminal study\cite{Uemura1989} established a scaling relation $T_c \propto \rho_s$ between $T_c$ and the superfluid density ($\rho_s$), while Homes et al.\cite{Homes2004} linked $T_c$ to the ratio of $\rho_s$ and normal-state DC conductivity ($\sigma_{\text{dc}}$), expressed as $T_c \propto \rho_s/\sigma_{\text{dc}}$. These empirical relations advanced the field by highlighting, for instance, the pivotal role of phase fluctuations \cite{Emery1995}. However, existing scaling laws lack clear characteristics of electron correlation quantities and universality across different supercritical families. The key is that they were unable to determine a unified energy scale similar to the Debye temperature in BCS theory\cite{McMillan1968}. This gap hinders the universal description of $T_c$ and masks the microscopic origins of UcS pairing.
    
     To address this issue, based on experimental observations and analysis of material structure, we establish a universal scaling relation: 
     \begin{eqnarray}
         k_{B}T_{c}^\star\cdot N_{\text{CP}}&=&\alpha \cdot U,
   	     \label{Eq_Ncp}
     \end{eqnarray} 
     where $k_{B}$ is the Boltzmann constant and $\alpha = 1/(16\pi$), $U$ is the on-site Coulomb repulsion, $N_{\text{CP}}$ (defined in subsequent section) quantifies the real-space extent of Cooper pairs (i.e., geometric constrains involving experimentally measurable coherence length $\xi_0$, lattice parameter $a$, and dimension $D$). This relation in Eq.~(\ref{Eq_Ncp}) has been exhaustively checked through meta-analysis of over 500 experimental datasets spanning 14 UcS families. It holds for $\sim$90\% of known UcSs (173 compounds) across a four-order-of-magnitude $T_c$ range (0.08–133 K), while systematically excluding conventional phonon-mediated superconductors (these show a deviation of $\gtrsim$80\% from the relation). The invariance of $\alpha$ across disparate material family indicates a unified Coulomb-driven pairing mechanism, where $N_{\text{CP}}$ spatially confines Cooper pairs.
    
\section{Data and Methods}

    The datasets used to validate universal scaling relation containing over 187 from 14 different material categories involved in over 500 experiments, including those of cuprates, iron-based superconductors, heavy fermion systems, organic charge transfer salts, Chevrel phases, alkali doped $C_{60}$, quasi-1D superconductors, twisted van der Waals materials (e.g., graphene, WSe$_2$, MoTe$_2$), nickelates, Kagome lattice materials, half-Heusler compounds, cobaltates, layered nitrides, and rhombohedral multilayer graphene. For detailed data, please refer to Experimental collection datasets in the Supporting Information (SI). For comparative analysis, representative BCS superconductors such as MgB$_{2}$\cite{Souma2003} and H$_{3}$S\cite{Drozdov2025} were listed as benchmarks. Although the classification of certain systems (e.g., Chevrel phases\cite{Pena2015}) as UcS or BCS-aligned remains debated, this ambiguity was systematically addressed through uniform methodological protocols implemented consistently across all materials in our study, minimizing selection bias in material selection.
    
    Methodologically, we adopted a standardized protocol (detailed in Appendix~\ref{app:A}) to extract five key experimental parameters: (\romannumeral1) zero-field maximum transition temperatures $T_{c}^{\star}$; (\romannumeral2) zero-temperature coherence lengths $\xi_{0}$; (\romannumeral3) crystallographic descriptors (crystal structures and lattice constants $a, b, c$); (\romannumeral4) superconducting dimensionality $D$; and (\romannumeral5) on-site Coulomb interactions $U$. Complete data are presented in Tables S\uppercase\expandafter{\romannumeral3}-S\uppercase\expandafter{\romannumeral37} of the SI. These parameters constitute a minimal universal set capturing the essential characteristics of UcSs. Among these five parameters, the first three are direct experimental observables. In contrast, $D$ and $U$ require indirect determination using a combination of transport measurements, spectroscopic techniques (e.g., ARPES, XPS, EPS, $\mu$SR), and computational modeling (e.g., DFT+U, cRPA). While $U$ is theoretically defined as the intra-atomic $d/f$-electron repulsion, its extraction exhibits method-dependent variability; inter-technique discrepancies could even lead to deviations of up to about $\pm30\%$ deviations in specific systems. Nevertheless, as demonstrated later, the universal relationship identified in this work remains robust with high precision even under random Monte Carlo sampling within the error bounds of $U$ and $N_{\text{CP}}$ (see Appendix~\ref{app:B}).
    
    Physically, to uniformly characterize the spatial extent of Cooper pairs across disparate materials, we introduce a dimensionless parameter $N_{\text{CP}}$. This parameter quantifies the number of equivalent correlated atomic sites covered by a Cooper pair in real space, based on the previously defined observables (\romannumeral2), (\romannumeral3), and (\romannumeral4). It is defined as $N_{\text{CP}}=\gamma\xi_{0}^{D}/\Omega_{\text{SC}}$, where $\Omega_{\text{SC}}$ denotes the minimal periodic superconducting unit within the unit cell (e.g., $\Omega_{\text{SC}}=ab$ corresponds to the 2D CuO$_2$ plane area in cuprates, and $\Omega_{\text{SC}}=abc$ refers to the 3D unit cell volume of fullerenes), and $\gamma$ is an integer that counts the equivalent correlated sites within $\Omega_{\text{SC}}$ (e.g., Cu atoms in CuO$_{2}$ planes, Fe-1 or Fe-2 sites in FeAs layers). Crucially, the ratio $\Omega_{\text{SC}}/\gamma$ is invariant to unit-cell definitions, owing to underlying crystal symmetries. For instance, in 45°-diagonal Cu-O configurations with $\gamma=2$, the symmetry-protected lattice expansion to $\sqrt{2}a$ preserves the invariance of $\gamma/\Omega_{\text{SC}}$.
    
\section{Universal Scaling of $T_c$}

    \begin{figure}[tb]
    	\centering
    	\includegraphics[scale=0.21]{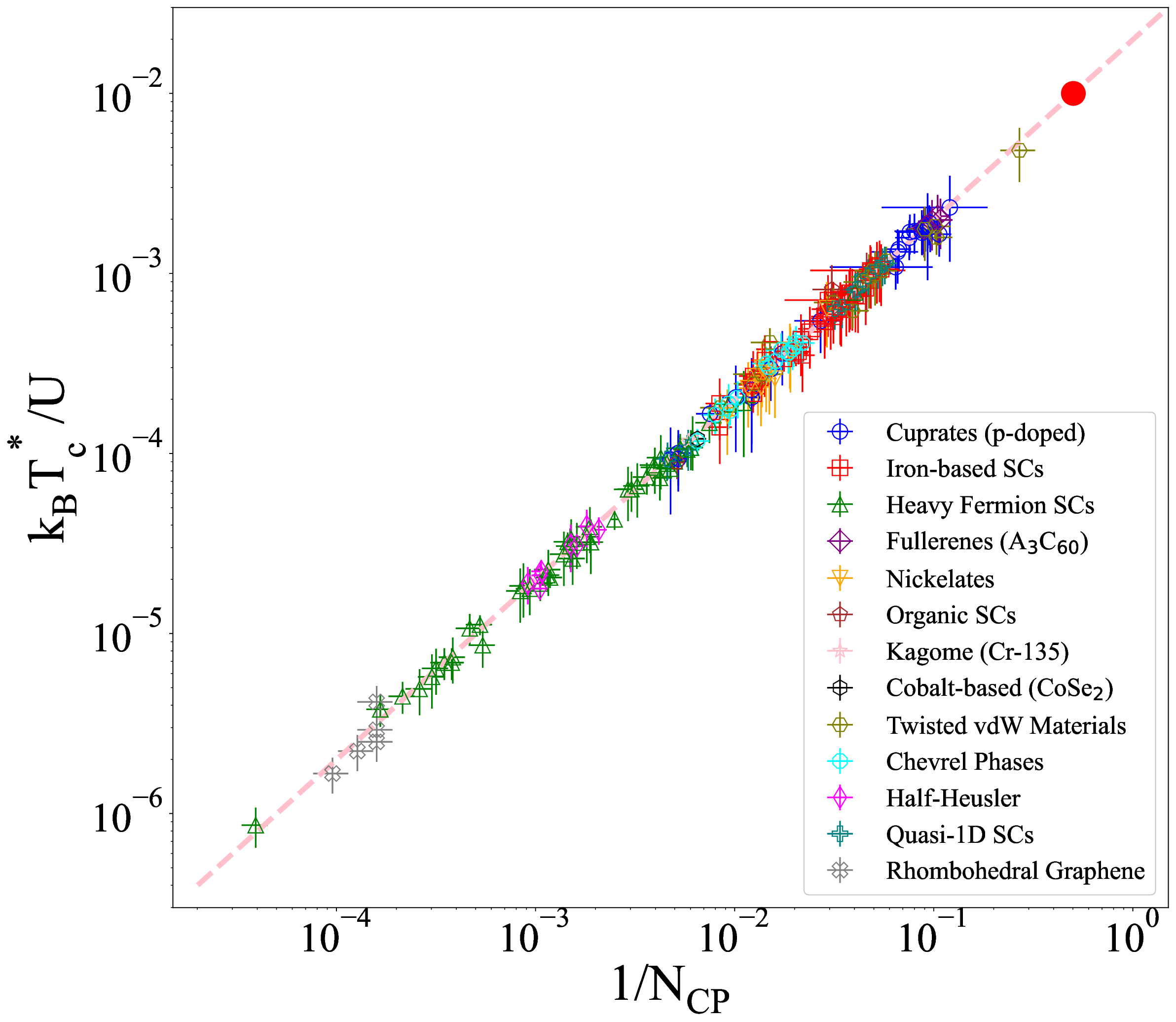} 
    	\caption{\textbf{Universal scaling of $T_{c}^\star$ in UcSs}. The normalized maximal critical temperature $T_{c}^\star/U$ shows a universal linear dependence on $1/N_{\text{CP}}$ across 160+ UcSs from 13 families involved in over 500 experiments. The solid red circle marks the limiting case at $N_{\text{CP}}=2$, giving $k_{\text{B}}T_{c}^\star/U\approx0.01$. The pink dashed line corresponds to Eq.~(\ref{Eq_Ncp}).}
    	\label{fig:1}
    \end{figure}

    The scale relation in Eq.~(\ref{Eq_Ncp}) clearly reflects the energy characteristic quantity relationship in UcSs. The left-hand side features $T_c$, which has the dimensions of energy and is generally proportional to the energy gap $\Delta_{\text{gap}}$. $N_{\text{CP}}$, a geometry-related parameter, characterizes the real-space confinement of Cooper pairs. The right-hand side contains $U$---a physical quantity representing electron correlation strength. The physical implication of this universal relation is that Cooper pair formation arises from correlation-driven effects under geometric confinement. 
    
    After the database was completed (with details provided in the SI), a log-log plot of log($k_{B}T_{c}/U$) versus log($1/N_{\text{CP}}$) was generated, as shown in Fig.~\ref{fig:1}. Fig.~\ref{fig:1} compiles data from 173 materials among the 187 compounds mentioned above (the other dozen or so will be specifically discussed as the suspected deviations in Fig.~\ref{fig:2}). These 173 materials, across 13 different families, include: hole-doped cuprates, iron-based superconductors, Ce-/U-/Pu-/Np-based heavy fermion systems, organic charge transfer salts ($\kappa$-/$\beta^{\prime\prime}$-(ET)$_{2}$X, Bechgaard salts), twisted-angle $n$-layer graphene (with layers $n=2\sim5$, filling factor $\nu\sim-2$), twisted WSe$_2$/MoTe$_2$, Jahn-Teller-active fullerenes Rb$_{x}$Cs$_{3-x}$C$_{60}$, quasi-one dimensional correlated superconductors, Chevrel phases, nickelates, the Kagome material (CsCr$_{3}$Sb$_{5}$), layered cobalt oxychalcogenides (-[CoSe$_{2}$]), half-Heusler compounds, and rhombohedral graphene. These data align closely with the pink dashed line in Fig.~\ref{fig:1}, given by Eq.~(\ref{Eq_Ncp}). It is shown that different superconducting material families correspond to a common constant slope, validating the existence of a universal scaling relationship represented by Eq.~(\ref{Eq_Ncp}). 
    
    It is worth noting that in order to evaluate the uncertainty of experimental measurements, Monte Carlo sampling (100,000 simulations) was performed on $U$ and $N_{\text{CP}}$ within their respective error ranges: the goodness of fit $R^2$ remained robust ($>$0.97), the $\alpha$ deviation was $<$3\%, and the parameter distributions followed normality (Table~\ref{tab:statistics_comparison}, Fig.~\ref{fig:AppendixB}), verifying the robustness of the scaling law. 
    
    The distribution of data points in Fig.~\ref{fig:1} exhibits distinct clustering: cuprates data points (blue circles) accumulate in the upper-right quadrant, heavy-fermion systems data points (green triangles) and rhombohedral graphene data points (gray crosses) congregate in the lower-left quadrant, and data points from other material families occupy intermediate regions---with all clusters aligning with specific segments of the scaling line (pink dashed line in Fig.~\ref{fig:1}). As detailed in Supplementary Section \uppercase\expandafter{\romannumeral2}-A, UcSs are dominated by transition-metal elements (Cu, Fe, Ni, Ce) with $d$/$f$-electrons, while carbon ($p$-electrons) in organic superconductors is the only non-transition element. 
    
    Interestingly, we can observe that all data points are distributed below the solid red circle on the universal linear line. The solid red circle correspond to the minimum $N_{\text{CP}}$. Thus, this universal scaling relationship suggests the existence of an upper limit for the transition temperature. For a typical $U = 2{-}6 \, \text{eV}$, the limit on the transition temperature is $T_{\text{c}}^\star \lesssim 231{-}692$ K, thereby leaving open the possibility of achieving room-temperature superconductivity. 
    
\section{Comparison with Uemura’s and Homes’ Laws}

    \begin{table*}[!tbh]
    	\centering
    	\caption{Display of comparison between theoretical values calculated based on Eq.~(\ref{Eq_Ncp}) and experimentally measured values from some conventional cases.}
    	\begin{tabular}{|l|l|l|l|l|}
    		\toprule
    		\rowcolor[gray]{0.94} 
    		Observable Quantities & Scaling Law & Technology & Results from Eq.~(\ref{Eq_Ncp}) & Experimental Values\\
    		\midrule
    		Uemura's Plot  & \multirow{3}[4]{*}{\tabincell{c}{\tabincell{l}{Uemura's \\(\textit{Lacking data}) }  }} & $\mu$SR &  $\rho_s\propto 1/N_{\text{CP}}\propto T_c$  & $\rho_s\propto T_c$ \\
    		\cmidrule(r){1-1}\cmidrule(l){3-5}  
    		Cooper-pair per Unit Cell (LSCO) &  & MI & 0.12$\pm$0.06 & $\sim$0.137 \\
    		\cmidrule(r){1-1}\cmidrule(l){3-5}  
    		$T_c^*$ (twisted graphene, $\nu\sim-2$) &  & RFRC & 1.2$\pm$0.25 K & $\sim$1.1 K \\
    		\midrule
    		$\lambda^2_L(0)$ of BaFe$_2$(As$_{1-x}$P$_{x}$)$_2$ at $x=0.3$ & \multirow{2}[2]{*}{\tabincell{c}{anti-Uemura's \\ (\textit{Inconsistent})}} & $\mu$SR & divergence & divergence \\
    		\cmidrule(r){1-1}\cmidrule(l){3-5} 
    		$T_c^*$ of BaFe$_2$(As$_{1-x}$P$_{x}$)$_2$ at $x=0.3$ &  & Transport &  31.7 K (see SI Table S\MakeUppercase{\romannumeral10}) & 29.2 K \\
    		\midrule
    		$T_c^*$ (unstrained Sr$_2$RuO$_4$) &  \multirow{4}[8]{*}{\tabincell{c}{Beyond Homes'\\ (\textit{Inconsistent})}} & \multirow{4}[8]{*}{\tabincell{c}{Transport}}  & 1.6$\pm$0.4 K & 1.5 K \\
    		\cmidrule(r){1-1}\cmidrule(l){4-5}  
    		$H_{c2}/T_c^2$ (unstrained Sr$_2$RuO$_4$) &    &   &  0.0286$\pm$0.0072 T/K$^2$ & 0.029 T/K$^2$ \\
    		\cmidrule(r){1-1}\cmidrule(l){4-5}  
    		$T_c^*$ (strained Sr$_2$RuO$_4$) &   &   &  3.6$\pm$1.2 K & 3.4 K \\
    		\cmidrule(r){1-1}\cmidrule(l){4-5}  
    		$H_{c2}^{\parallel b}/T_c$ (strained Sr$_2$RuO$_4$) &   &   &  \tabincell{l}{0.96$\pm$0.32 T/K\\(zero-temperature)} & \tabincell{l}{$\sim$1 T/K at 20 mK \\ ($d\chi/dB|_{\text{max}}$ criterion)} \\
    		\bottomrule
    	\end{tabular}%
    	\label{tab:cross}%
    \end{table*}
   
    The universal scaling relation can naturally incorporate and extend Uemura's and Homes' relations. For Uemura’s relation\cite{Uemura1989}, the intrinsic correlation between $N_{\text{CP}}$ and superconducting carrier density $n_s$---where $n_s$ is derived from zero-temperature London penetration depth $\lambda(0)$ (measured via $\mu$SR, microwave resonance, or magnetic levitation)---establishes a microscopic basis for this relation. Given that Cooper pairs exhibit negligible spatial overlap\cite{Coffman2000,Chen2024}, the relation $n_s \Omega_{\text{u.c.}} \propto 1/N_{\text{CP}} = k_B T_c^\star/(\alpha U)$ (where $\Omega_{\text{u.c.}}$ denotes the unit cell volume) directly accounts for both Uemura’s scaling ($\rho_s \propto T_c$, with $\rho_s = n_s/m^\star = 1/(\mu_0 e^2 \lambda(0)^2)$ representing the superfluid stiffness) and Božović’s scaling\cite{Bozovic2016} (measured via mutual inductance, i.e., MI). Specifically, under constant $m^*$ and $U$, formula in Eq.~(\ref{Eq_Ncp}) reduces to Uemura’s law, while extending it to scenarios beyond Uemura’s applicability---such as the anti-correlation between $T_c$ and $\rho_s$ in the iron-based superconductor BaFe$_2$(As$_{1-x}$P$_{x}$)$_2$ in the vicinity of quantum critical point\cite{Hashimoto2012}, where $\rho_s \propto T_c/m^*$ reaches a minimum even as $T_c$ peaks. For Božović’s LSCO samples, this yields $n_s \Omega_{\text{u.c.}} = 4k_B T_c/(\alpha U) \sim 0.12 \pm 0.06$ per unit cell (factor 4 accounting for bilayer CuO$_2$ and paired electrons), which consists with the reported value of $\sim 0.137$\cite{Zaanen2016} for $m^* = m_e$.
    
    In addition, a more explanatory verification comes from the twisted three-layer graphene, where superfluid stiffness measured using radio-frequency resonant circuits (RFRC)\cite{Banerjee2025} corroborates our earlier discussion. Specifically, $n_s$-$N_{\text{CP}}$ relation yields the expression for $T_c$: $k_{B}T_{c} = \alpha U \cdot 4k_{B}m^{*} \cdot \left( 3\sqrt{3}a_{M}^{2}/2 \right) (w/l) \tilde{\rho}_{s}/(2\hbar^{2})$, where $\tilde{\rho}_{s}$ denotes the superfluid stiffness measured in the cited work, $w$ and $l$ are the device width and length ($w/l\sim1/5$), $a_{M}$ is the Moire period, and $m^{*}=m_e$. Substituting data near the filling factor $\nu\sim-2$ yields transition temperatures consistent with experimental measurements (see more details in SI Section \uppercase\expandafter{\romannumeral2}-D), encompassing a wide range around maximal $T_c$, here $T_c^*\approx1.2\pm 0.25$ K matches the experimental value of $\sim$1.1 K.
    
    For Homes’ law\cite{Homes2004} ($\rho_s \propto \sigma_{\text{dc}} T_c$), limited understanding of $\sigma_{\text{dc}}$ precludes deriving a quantitative relationship. However, our universal relation encompasses not only unconventional superconductors (UcSs) that satisfy Homes’ law but also explains notable exceptions such as Sr$_{2}$RuO$_{4}$\cite{Dordevic2013}. Based on analysis presented in Supplementary Section \uppercase\expandafter{\romannumeral5}, the universal relation yields theoretical values of $T_{c}^\star=1.6$ K and $ H_{c2}/T_{c}^{2}=\Phi_{0}k_{B}^{2}/[2\pi\left(\alpha Ua\right)^{2}] \sim 0.0286\pm0.0072$ T/K$^2$ for unstrained Sr$_{2}$RuO$_{4}$, which show excellent agreement with experimental measurements ($T_{c}^\star\sim1.5$ K\cite{Kittaka2009} and $H_{c2}/T_{c}^2=0.029$ T/K$^2$ reported by Mao et al.\cite{Mao1999}). Once uniaxial pressure is applied along the $\left\langle100 \right\rangle$ direction\cite{Steppke2017}, $T_c$ can be enhanced to 3.4 K, with $H_{c2}^{\parallel b}/T_c$ ($\approx$1 T/K at 20 mK by the $d\chi/dB|_{\text{max}}$ criterion) remaining constant\cite{Jerzembeck2023,Pustogow2019}---rather than $H_{c2}^{\parallel c}/T_c^2$. According to detailed analyses in the SI, our universal scaling relation yields $T_c = 3.6\pm1.2$ K and $H_{c2}^{\parallel b}/T_c=0.96\pm0.32$ T/K.
    
    Systematic comparisons between theoretical calculations from our relation with experimental data across techniques are provided in Table~\ref{tab:cross}, where we categorize the data into three groups corresponding to Uemura’s and Homes’ laws---encompassing both data that qualitatively satisfy these laws and those that do not---and find a quantitatively  consistent agreement across all cases.

\section{Deviant Cases and the Conventional-Unconventional Boundary}

    \begin{figure}[tb]
    	\centering
    	\includegraphics[scale=1.15]{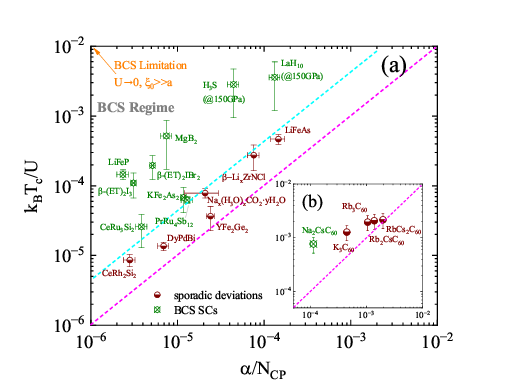} 
    	\caption{\textbf{Distinctive scaling regimes of UcSs versus BCS superconductors.} (a) Superconductors that diverge from the universal scaling relation in Eq.~(\ref{Eq_Ncp}) are distinctly color-coded. The cyan dashed boundary separates BCS superconductors (green circles) from UcSs. The intermediate parameter space (between cyan demarcation and purple universal line in Fig.~\ref{fig:1}) hosts transitional UcS candidates. (b) The deviation scaling relationship of A$_{3}$C$_{60}$ under different degrees of alkali metal intercalation is displayed..}
    	\label{fig:2}
    \end{figure}    
        
    Fig.~\ref{fig:1} presents results of 173 materials from 13 families, among 187 compounds across 500 experiments. The study found that those a dozen materials not included in Fig.~\ref{fig:1} exhibit limit deviations from the linear relationship. We have marked the outliers in Fig.~\ref{fig:2}(a), situated in the range between the cyan and pink lines in Fig.~\ref{fig:2}(a). These materials that deviate to a certain extent from Eq.~(\ref{Eq_Ncp}) are selected A$_{3}$C$_{60}$ polymorphs (A=Na, K, Rb), cobalt oxide hydrates (Na$_{x}$(H$_{3}$O)$_{z}$CoO$_{2}$ · yH$_{2}$O), along with isolated outliers such as DyPdBi, CeRh$_{2}$Si$_{2}$, LiFeAs and YFe$_{2}$Ge$_{2}$, the layered nitride family is also excluded from Fig.~\ref{fig:1}. Relevant data and references are also provided in the tables included in the SI. As shown in Fig.~\ref{fig:2}(a), UcSs exhibit limited deviations (magenta half-filled circles), clustering above the universal purple dotted line within error margins---this indicated $T_c^\star$ acts as a lower bound for $T_c$ at fixed $N_{\text{CP}}$. 
    
      \begin{table*}[htbp]
    	\centering
    	\caption{Values represent the average of relevant parameters and the extent of deviation from the scaling relation for various materials. Detailed references are provided in the SI.}
    	\begin{tabular}{|l|l|c|c|c|c|c|c|c|c|c|c|c|c|}
    		\toprule
    		Family & Material  & $D$   & Crystal &$\gamma$ & $a/\mathring{\mathtt{A}}$ & $b/\mathring{\mathtt{A}}$ & $c/\mathring{\mathtt{A}}$ & $\xi{_{ab}}/\mathring{\mathtt{A}}$ & $\xi_c/\mathring{\mathtt{A}}$ & $U$/eV  &  $T_{c}^\star$/K &  Eq.~(\ref{Eq_Ncp})/K & Deviation \\
    		\midrule
    		\multirow{3}[6]{*}{\tabincell{l}{n-doped Cuprates} } & PCCO & \multirow{3}[6]{*}{2} & \multirow{3}[6]{*}{OR} & \multirow{3}[6]{*}{1} & 3.9615 & 3.9615 & 12.214 & 60.8  &   -    &   -    &    19.5   &   -    &  - \\
    		\cmidrule{2-2}\cmidrule{6-14}          & NCCO &       &       &       & 3.956 & 3.956 & 12.108 & 70.1  & 3.4   &   -   &   20.5    &   -    & -  \\
    		\cmidrule{2-2}\cmidrule{6-14}          & SLCO  &       &       &       & 3.988 & 3.988 & 3.397 &   -    &   -    &   -    &  -     &   -    & -  \\
    		\midrule
    		\multirow{4}[8]{*}{\tabincell{l}{A$_3$C$_{60}$ $\ddag$ \\(A=Na,K,Rb)}} & RbCs$_{2}$C$_{60}$ & \multirow{4}[8]{*}{3} & \multirow{4}[8]{*}{FCC} & \multirow{4}[8]{*}{4} &         14.48    &   14.48    &   14.48    &   19.68    &    19.68     & \multirow{4}[8]{*}{1.3}  &   32.5    &    29.9   &  8\% \\
    		\cmidrule{2-2}\cmidrule{6-10} \cmidrule{12-14}          & Rb$_2$CsC$_{60}$ &       &       &       &   14.36    &   14.36    &   14.36    &    22    &   22   &       &    31.4   &    20.9   &  33\%\\
    		\cmidrule{2-2}\cmidrule{6-10} \cmidrule{12-14}         & Rb$_3$C$_{60}$ &       &       &       &    14.33   &    14.33   &   14.33    &   24.03    &   24.03     &       &   29.4    &   15.9    &  46\%\\
    		\cmidrule{2-2}\cmidrule{6-10} \cmidrule{12-14}         & K$_3$C$_{60}$ &       &       &       &   14.16    &   14.16    &  14.16     &  31.58   &   31.58    &       &  19.2     &    6.8   & 65\% \\
    		\midrule
    		\multirow{2}[4]{*}{\tabincell{l}{Layered Nitrides}} & $\beta$-Li$_x$ZrNCl & 2 &  \multirow{2}[4]{*}{RB} & 1     & 3.59 & 3.59 &   -    & 54.1  &    -   &  5   &   15.9    &   4.4    & 72\% \\
    		\cmidrule{2-3}\cmidrule{5-14}      & $\alpha$-Na$_{0.16}$TiNCl &   -    &       & -     & 4.0187 & 3.2738 & 16.884 & 33    & 28    &   -    &   18    &   -    & - \\
    		\midrule
    		Cobaltates & \tabincell{l}{Na$_{x}$(H$_{3}O$)$_{z}$-\\-CoO$_{2}\cdot y$H$_{2}$O} & 2     &   RB    &   2    &   5.6584    &   5.6584    &    -   &  115   &   -    &  4.5     &     4.1    &   1.09    &  73\%  \\
    		\midrule
    		Iron-based SC & LiFeAs  $\ddag$ & 2     & TG & 1     & 3.77  & 3.77  & 6.34  & 44 & 22.5 & 3.14  &   17    &   5.3    & 69\% \\
    		\midrule
    		\multirow{2}[4]{*}{Heavy Fermion} & CeRh$_2$Si$_2$ & 2     & TG & 1     & 4.083 & 4.083 & 10.171 & 343   &   -    & 5 &    0.5   &    0.16   &  68\% \\
    		\cmidrule{2-14}          & YFe$_2$Ge$_2$ & 2     & TG & 1     & 3.964 & 3.964 & 10.457 &   114    & - & 4 &  1.7    &   1.12    &  34\% \\
    		\midrule
    		Half Heusler & DyPdBi & 2     & FCC & 2     & 6.6343 & 6.6343 & 6.6343 & 251.58 & 251.58 & 5.8  &   0.92  &   0.47    &  49\% \\
    		\bottomrule
    	\end{tabular}%
    	
    	\smallskip
    	\footnotesize
    	\textbf{Note:} The dash ('–') denotes currently unavailable data; the symbol '$\ddag$' marks materials with disputed classification as UcS.  \\
    	\textbf{Abbreviations:} OR: Orthorhombic; FCC: Face-centered cubic; RB: Rhombohedral; TG: Tetragonal.
    	\label{tab:dev}%
    \end{table*}%
    
    The experimental data of n-doped UcSs [such as Pr$_{2-x}$Ce$_{x}$CuO$_{4-\delta}$(PCCO), Nd$_{2-x}$Ce$_{x}$CuO$_{4-\delta}$(NCCO) and Sr$_{1-x}$La$_{x}$CuO$_{2}$(SLCO)] and $\alpha$-form layered nitrides are not included both in Fig.~\ref{fig:1} and \ref{fig:2} due to incomplete experimental parameters. As shown in Table.~\ref{tab:dev}, definitive values for $T_c^*$ and the interaction strength are unavailable. The reported $H_{c2}$ data sometimes correspond to a generic $T_c$, generally below the maximum transition temperature, rather than the $T_c^*$ defined by Eq.~(\ref{Eq_Ncp}). In addition, since there is no reference value for $U$, it is obviously incorrect to use $U$ under hole doping to estimate $n$-doping. The data under this substitution will inevitably deviate from the universal scaling law shown in Eq.~(\ref{Eq_Ncp}), and the resulting deviation cannot be used as a basis for not conforming to the Eq.~(\ref{Eq_Ncp}) relationship. Because UcSs do not have particle-hole symmetry\cite{Anderson2006}, the interaction energy under electron doping is different from that under hole doping. Adding an electron to a hole typing is asymmetric when lacking an electron\cite{Pan2000}. Electron doping and hole doping are completely different\cite{Segawa2010}. Therefore, the verification of UcSs under $n$-doping requires more comprehensive experimental data support.
    
    Fig.~\ref{fig:2}(b) shows data points obtained from C$_{60}$ experiments, which correspond to different degrees of alkali metal intercalation. Similar to hole doping in copper oxides, the maximum intercalation level is achieved in the Rb/Cs intercalated cases, consistent with the scaling relation of Eq.~(\ref{Eq_Ncp}). However, the data in Fig.~\ref{fig:2}(b) do not correspond to estimates made using $T_c^\star$; they show a similar trend to the deviation observed in cuprates with varying hole concentrations. Therefore, the deviation caused by not taking the maximum conversion temperature $T_c^\star$ value does not indicate that the actual situation contradicts the scaling relationship shown in Eq.~(\ref{Eq_Ncp}).
          
    Other observed UcS deviation in UcSs systematically fall between the universal (purple dashed) and empirical (cyan dashed) lines in Fig.~\ref{fig:2}(a), showing qualitative agreement with Eq.~(\ref{Eq_Ncp}). By contrast, conventional BCS superconductors (green crossed circles) show deviations beyond this limit, clustering in the top-left regime of Fig.~\ref{fig:2}. This segregation—with BCS systems (e.g., Hg, Al; $U\rightarrow0$, $\xi(0)\gg a$) distinct from UcS-dominated regions---validates the boundary’s role in distinguishing pairing mechanisms: phonon-mediated attraction (BCS) versus Coulomb-driven correlations (UcSs).
    
    It is worth noting that a blurred boundary exists between the two regimes; this boundary divides the entire region into conventional and unconventional parts. Materials near this boundary have sparked debates over their classification as BCS or unconventional. We approximately located this boundary using several materials that straddle the conventional and unconventional regimes. In A$_3$C$_{60}$, for instance, the level of alkali-metal intercalation tunes the ground state: Rb$_{3-x}$Cs$_{x}$C$_{60}$ falls inside the unconventional domain, whereas the more deviant Na$_2$CsC$_{60}$ lies outside the cyan boundary. This crossover is from a correlated superconductor to BCS one (Supplementary Section \uppercase\expandafter{\romannumeral2}-C). Similar crossovers occur in $\beta$-Li$_x$ZrNCl\cite{Nakagawa2021}. The cyan line is mainly drawn according to those crossover systems.

\section{Discussions and Outlook}
     
    In experimental practice, a key utility of the scaling relation lies in supporting experimental investigations, allowing direct estimation of $T_{c}^\star$ or $H_{c2}$ for new or understudied superconducting materials. For example, applying Eq.~(\ref{Eq_Ncp}) to recent ambient-pressure (La,Pr)$_{3}$Ni$_{2}$O$_{7}$ thin-film\cite{Zhou2025} yields $T_{c}^\star\approx29\pm10$ K (90\%$\rho_{n}$ criterion) using: (\romannumeral1) quasi-2D superconducting behavior (three-unit-cell thickness, consistent with the 2D-Tinkham model), (\romannumeral2) tetragonal phase ($\gamma=1$) with $a\approx3.75$ \AA~lattice constant matching the SrLaAlO$_{4}$ substrate, (\romannumeral3) in-plane coherence length $\xi_{ab}(0)=2.2$ nm (also from the 90\%$\rho_{n}$ criterion), and (\romannumeral4) on-site interaction $U=4.3\pm1.5$ eV, consistent with other nickelate superconductors. This prediction agrees with experimental uncertainty (34 K, 90\%$\rho_{n}$ criterion), with a 15\% deviation primarily attributed to two factors: (1) vertical confinement enhancing $U$ by $\sim$13\%, and (2) systematic errors ($\sim$2\%) in extrapolating $H_{c2}(0)$ and lattice parameter determination. Similarly, if $T_c^\star$ is known, we can conversely predict $H_{c2}$. For instance, for the material system (Ca$_x$La$_{1-x}$)(Ba$_{1.75-x}$La$_{0.25+x}$)Cu$_3$O$_y$ (CLBCO,  $a=3.872$ \AA, $T_c^{\star}$ $\sim$ 81.5 K when $x = 0.45$)\cite{Chabaud1996}, whose experimental $H_{c2}(0)$ data is lacking, we predict its upper critical field $H_{c2}(0) \sim 130$ T.
    
    In addition, Eq.~(\ref{Eq_Ncp}) provides a feasible reference for modulating and preparing high-temperature superconductors by enhancing Coulomb interactions through lattice engineering and optimizing the spatial constraints on Cooper pairs.

    From a theoretical perspective, the fact that seemingly distinct material families follows the same fundamental scaling laws raises the theoretical exploration question of whether different UcS families share mechanism. This thereby provides clues for developing a quantitative framework to test hypotheses and elucidate the underlying pairing mechanism. In addition, for theoretical researchers, two priority directions emerge from this work: (1) Resolving deviations in electron-doped cuprates and other systems to rigorously define the boundary between conventional and unconventional superconductivity. (2) Based on the prompt provided by Eq.~(\ref{Eq_Ncp}), offer a reasonable theoretical explanation for the possible highest $T_c^\star$ of various superconducting parent materials through their dimensional properties.

	\section*{Acknowledgments}
	
	We are particularly grateful for the financial support from NSFC12088101. This work also received funding support from CPSF2024M752758, MOST2022YFA1402701, and NSF11774006.
	
	We thank Chao Cao, Guang-Han Cao, Xiao-Jia Chen, Lun-Hui Hu, Zhi-Bin Huang, Yi-Na Huang, Lin Jiao, Amit Keren, Chang-Xiao Li, Hui Li, Wen-Yuan Liu, Yang Liu, Ming Shi, Zhen-Tao Wang, Yan-Wu Xie, Zhu-An Xu, Sheng Yang, Huai-Yang Yuan, Hui-Qiu Yuan, and Ya-Nan Zhang for their constructive discussions on experiments and numerical simulations with us.





\appendix

\section{Method: Standardized Protocols}
\label{app:A}

(\romannumeral1) $T_{c}^{\star}$: Defined as the doping- and pressure-insensitive maximum transition temperature. It signifies the optimal superconducting order performance and often coincides with maximal deviation from conventional mechanisms. $T_{c}^{\star}$ is determined through transport measurements using either onset criterion, 90\% normal-state resistance ($\rho_{n}$) recovery, or a 50\% midpoint criterion when 90\% recovery data is unavailable. The dominant error source stems from statistical uncertainty ($\lesssim$ 5\%) across multiple experimental realizations under identical doping/pressure conditions.

(\romannumeral2) $\xi_{0}$: Calculated via the Ginzburg-Landau relation $\xi(0)=\sqrt{\Phi_{0}/(2\pi H_{c2}\left(0\right))}$, where $\Phi_{0}\approx2.0678\times10^{-15}\text{Wb}$ is the flux quantum, $H_{c2}$ is determined from transport measurements under $T_{c}^{\star}$-consistent protocols. Potential errors arise from extrapolations (e.g., Werthamer-Helfand-Hohenberg (WHH) formula\cite{Werthamer1966}, empirical relations\cite{Zang2023, Wei2025}) to zero temperature when experimental field or temperature ranges are limited. The optimal solution exhibiting the best fitting performance among documented methodologies in the literature was selected for implementation.

(\romannumeral3) Crystal structures and lattice constants $a,b,c$: The database incorporates low-temperature crystallographic data obtained through synchrotron X-ray diffraction with Rietveld refinement, prioritizing cryogenic structural characterization to circumvent temperature-induced phase transitions. Lattice parameter determinations exhibit $\sim$0.5\% systematic uncertainty, with Bravais lattice types classified according to International Crystallographic Tables conventions. This protocol ensures atomic-scale accuracy in structural descriptors critical for superconducting pairing analysis.

(\romannumeral4) $D$: Determined through a consensus of following complementary criteria: crystallographic anisotropy, Fermi-surface topology via ARPES, anisotropic Coherence Lengths, cross-validation with transport properties or thermodynamic responses. e.g., Cuprates ($D=2$, CuO$_{2}$-plane), Bechgaard salts ($D=1$, via Fermi-surface topology and hopping anisotropy), UBe$_{13}$ ($D=3$, cubic symmetry with isotropic $H_{c2}$).

(\romannumeral5) $U$: Given the experimental challenges in quantifying electron-electron interactions, we adopt the dominant on-site contribution to define the Hubbard parameter $U$ through the Coulomb energy cost formalism\cite{Anisimov1991}: $U = E(e^{n+1}) + E(e^{n-1}) - 2E(e^{n})$, where $E_{n}$ denotes the total energy of the $d/f$-electron system with $n$ electrons. In practice, $U$ values are determined as averages from spectroscopic measurements (e.g., angle resolved photoemission spectroscopy (ARPES), extended photoelectron spectrum (EPS), auger spectrum, core-level spectra in X-ray photoelectron spectroscopy (XPS)) if available, whereas for systems lacking spectroscopic data, we employ first-principles calculations averaging over different approximations. Multi-orbital corrections are systematically incorporated via the Dudarev scheme\cite{Dudarev1998} $U_{\text{eff}}=U-J_{\text{H}}$, where the Hund's coupling $J_{\text{H}}$	accounts for the orbital exchange interaction. Practically, we exploit empirical evidence that $U$ exhibits minimal compositional variation ($\lesssim$10\%) across isoelectronic systems, e.g., isovalent substitutions in Hg-12(n-1)n, Bi-based cuprates, iron-based 122 or Fe-11 systems. This uniformity enables the assignment of class-specific $U$ parameters to major subfamilies.

For those the spectral data missed, there are several methods to determine the $U$ value in the literature: (1) Empirical Matching\cite{Antonides1977,Sawatzky1984}: This method involves estimating the $U$ value by matching the theoretically calculated band gap or lattice constant with the experimentally observed values. It is straightforward and often used to adjust the $U$ parameter in DFT+U calculations\cite{Anisimov1991} to more accurately reproduce experimental data. This approach is commonly referred to as the empirical method. (2) High-Precision Theoretical Methods: The band gap calculated using higher-precision methods, such as hybrid functionals\cite{Silva2007} (e.g., Heyd-Scuseria-Ernzerhof) or GW approximations, serves as a reference. The $U$ value in DFT+U is then adjusted to align with these high-precision results. (3) Linear Response Method: Commonly used in the VASP software package, this method was proposed by Cococcioni et al.\cite{Cococcioni2005} It involves redistributing the d/f orbital charge density (occupation number) of an atom by applying different response potentials (effective Coulomb and exchange interaction terms, LDAUU, LDAUJ). The $U$ value can also be obtained by calculating the effective Coulomb and the first partial derivative (linear response coefficient) of the exchange interaction term for the d/f orbital charge occupancy. (4) Calibration Method: Used in the Materials Project, this method calibrates the $U$ value by fitting the binary formation enthalpy from experiments. Least squares are employed to find the $U$ value that minimizes the error in REDOX reactions\cite{Zhou2004}. For instance, in the vanadium oxide system, different formation energy reactions are considered to identify the $U$ value range where the formation energy error crosses zero, and a linear equation is fitted to determine the $U$ value. (5) Unrestricted Hartree-Fock (UHF) Method\cite{Bach1994}: The UHF method begins with an unrestricted Hartree-Fock calculation, allowing electrons with different spins to occupy different molecular orbitals, thereby breaking spin symmetry. This broken symmetry solution introduces electron correlations beyond the standard Hartree-Fock method. Bulk materials are typically simulated using UHF calculations on finite-size clusters, where $U$ tends to converge as the cluster size increases. (6) Constrained Random-Phase Approximation (cRPA)\cite{Aryasetiawan2004,Vaugier2012}: Integrated into density functional codes based on the linearized augmented plane wave (LAPW) framework\cite{Madsen2005}, cRPA determines the Hubbard $U$ value by calculating the system's response function, avoiding the use of empirical values. The process starts by selecting a valid low-energy Hilbert space, constructing a partially shielded interaction within this space, and defining a Hubbard $U$ by equating the fully shielded interaction to the physical value.

In cuprates, the dominant role of the Cu$^{2+}$ $d_{x^2+y^2}$ orbital emerges from its hybridization with O$^{2-}$ $p$-orbitals, forming the basis for describing low-energy physics in these materials. Within the single-orbital framework, the Hubbard parameter $U$ relates to the superexchange interaction $J_{\text{ex}}$ through $U=4t^2/J_{\text{ex}}$, where $t$ denotes the hopping integral. This inverse proportionality between $U$ and $J_{\text{ex}}$ implies that $T_{c}\propto t^2a^2/J_{\text{ex}}$, providing a natural explanation for the enhanced $T_{c}$ observed in YBCO cuprates compared to LSCO systems, despite their weaker superexchange interactions (i.e., $J^{\text{YBCO}}_{\text{ex}}\sim110$ meV, $J^{\text{LSCO}}_{\text{ex}}\sim135$ meV\cite{Keimer1992}. $J_{\text{ex}}\approx0.12$ eV deduce $U\approx5.76$ eV with $t\approx0.4$ eV\cite{Rosch2005} ). For that we still want to consider the multi-orbital effect, the Hund's coupling $J_{\text{H}}$ modifies the effective Hubbard parameter through the relation $U_{\text{eff}}=U-J_{\text{H}}$ in the Dudarev scheme\cite{Dudarev1998}. Recent high-throughput calculations\cite{Moore2024} yield $U_{\text{eff}}=7.59 \text{eV}-1.12 \text{eV}=6.47 \text{eV}$, showing quantitative agreement with the estimations (6$\pm$2 eV) from spectroscopic methods: (a) Angle-Resolved Photoemission Spectroscopy (ARPES)\cite{Shen1995,Onsten2007} 6–8 eV; (b) Extended Photoelectron Spectroscopy (EPS)\cite{Shen1987,Eskes1990,Hybertsen1989} 6–10 eV; (c) X-ray Photoemission Spectroscopy (XPS)\cite{Fujimori1987,Zaanen1986} 6 eV; (d) Auger Electron Spectroscopy\cite{Marel1988,Fuggle1988} 5 eV; (e) Combined Spectroscopic/Scattering Analysis\cite{Sheshadri2023} 4 eV.

In iron-based superconductors, while multi-orbital effects dominate the electronic structure, the remarkable similarity of on-site Coulomb interactions ($U\approx3\sim5$ eV) across 122/1111 families shifts research focus to Hund's coupling $J_{\text{H}}$. The multi-orbital Hubbard model expresses these parameters via Slater integrals\cite{Anisimov1997}: $U=F^{\left(0\right)}$, $J_{\text{H}}=\left(F^{\left(2\right)}+F^{\left(4\right)}\right)/14$, where $F^{\left(0\right)}$ governs the radial electron-electron interaction, while higher-order $F^{\left(2\right)}$ and $F^{\left(4\right)}$ determine the strength of Hund's coupling. Usually for transition metals, $J_{\text{H}}$ is 10\%$\sim$20\% of $U$, i.e. $J_{\text{H}}\approx0.1U\sim0.2U$. Bringing it into the universal relationship in the main text, $T_{c}$ is directly proportional to $J_{\text{H}}$. 

In heavy fermion systems, orbital hybridization necessitates consideration of Anderson impurity model parameters, where the Schrieffer-Wolff transformation\cite{Schrieffer1966} maps hybridization strength $V$\cite{Anderson1961} to effective Kondo coupling $J_{K}\approx V^2/U$. This scaling governs the exponential dependence of Kondo temperature $T_{K}\approx \exp(-1/\rho J_{K})$, where $\rho$ denotes conduction electron density of states. The effective mass enhancement $m^{\star}/m \propto 1/T_{K}$ directly manifests heavy fermion behavior. The historical development of heavy fermion research initially focused on the Kondo single-impurity model with the infinite-$U$ approximation, where $U$ was treated as a fixed background parameter. However, systematic studies across different superconducting families have revealed substantial variations in $U$, typically spanning 2$\sim$6 eV, significantly exceeding the value of $J_{K}$($\sim0.1$ eV). This empirical evidence establishes $U$ as the dominant energy scale in determining correlated electron behavior.

In summary, parameters such as $J_{\text{H}}$, $V$, etc., play a crucial role in renormalizing $U$. In the main text, particularly within the context of iron-based superconductors and heavy Fermi systems, the $U$ we refer to is the effective $U_{\text{eff}} = U - J$ after renormalization. For specific families of superconductors, the magnitude of $U$ is relatively similar, making the focus on $J_{\text{H}}$ or $V$ more insightful for uncovering richer physics. However, when considering the commonalities across different superconducting families, variations in $U$ become more prominent. For instance, in the expression $U_{\text{eff}} = U(1 - J_{\text{H}}/U)$, if $U$ remains relatively unchanged, the effect of $J_{\text{H}}$ is dominant. Conversely, if $U$ undergoes significant changes, the impact of $J_{\text{H}}$ can be regarded as a second-order effect. Essentially, $U$ lays the foundation for the ground state correlation, while $J_{\text{H}}$ introduces spin fluctuations on this basis.

\section{Fitting Method and Assessing the Impact of Uncertainties}
\label{app:B}

\begin{table*}[tbh]
	\centering
	\caption{Statistical Comparison of fitting Parameters via Monte Carlo Sampling. To ensure the reproducibility of Monte Carlo sampling results, we set np.random.seed(1) (for NumPy) in Python 3.10.11.}
	\begin{tabular}{|l|l|r|r|r|r|}
		\hline
		Sampling Scenario  &   Parameters      &         Mean    &        Std Dev    &   95\% CI Lower   &    95\% CI Upper \\ \hline
		\multirow{3}{*}{Uniform Sampling} & $c_1$ & 0.9978 & 0.0104 & 0.9774 & 1.0183 \\ \cline{2-6} 
		& $c_2$ & -1.7146 & 0.0418 & -1.7966 & -1.6327 \\ \cline{2-6} 
		& $R^2$ & 0.9729 & 0.0024 & 0.9681 & 0.9776 \\ \hline
		\multirow{3}{*}{\tabincell{c}{Normal Sampling \\($\sigma=1$)}} & $c_1$ & 0.9999 & 0.0038 & 0.9924 & 1.0075 \\ \cline{2-6} 
		& $c_2$ & -1.7069 & 0.0161 & -1.7384 & -1.6753 \\ \cline{2-6} 
		& $R^2$ & 0.9939 & 0.0006 & 0.9927 & 0.9951 \\ \hline
		\multirow{3}{*}{\tabincell{c}{Normal Sampling \\($\sigma=1.96$)}} & $c_1$ & 1.0006 & 0.002 & 0.9967 & 1.0044 \\ \cline{2-6} 
		& $c_2$ & -1.7046 & 0.0082 & -1.7206 & -1.6886 \\ \cline{2-6} 
		& $R^2$ & 0.9964 & 0.0002 & 0.9959 & 0.9969 \\ \hline
	\end{tabular}
	\label{tab:statistics_comparison}
\end{table*}

\begin{figure}[tb]
	\centering
	\includegraphics[scale=0.11]{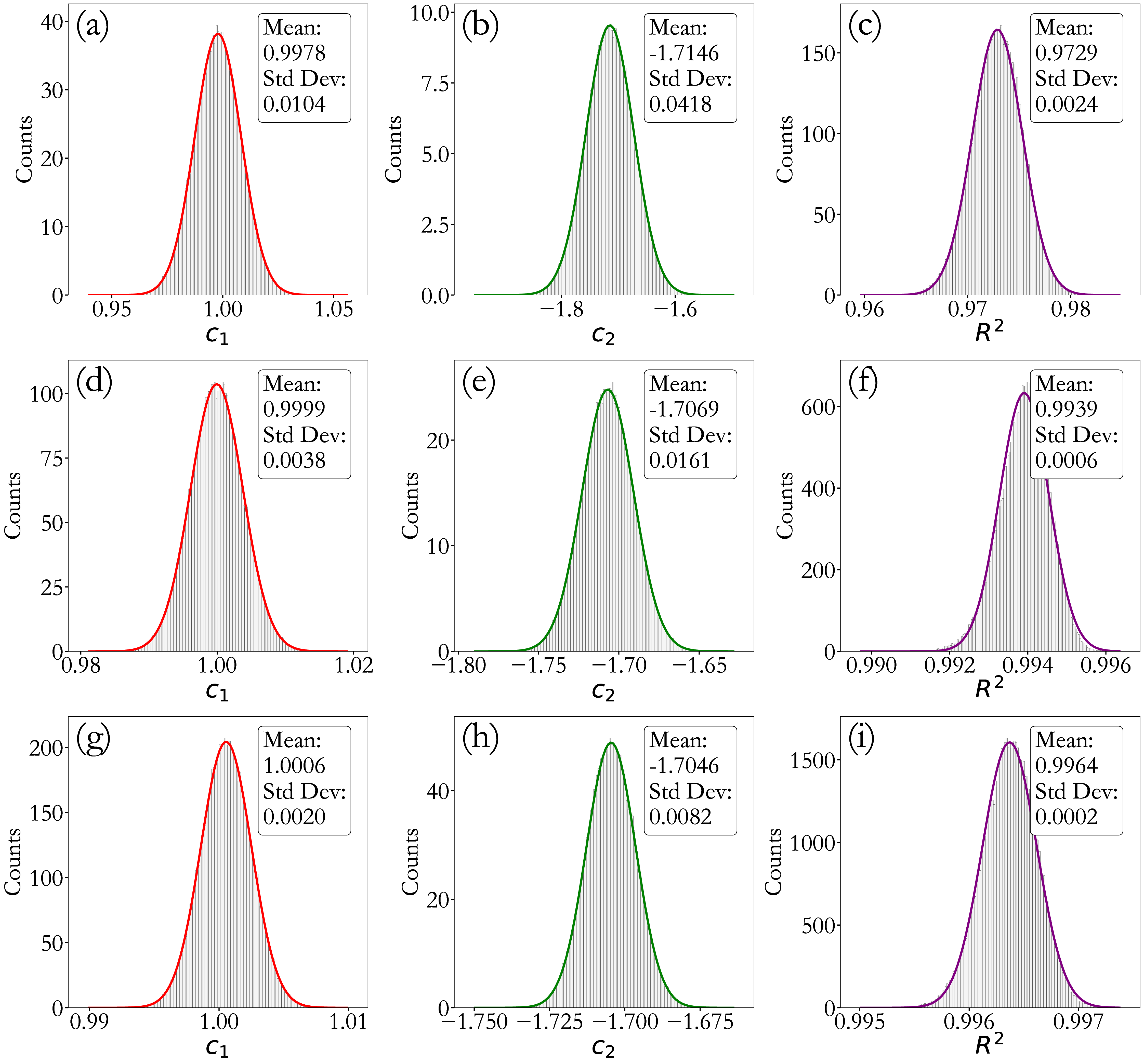} 
	\caption{This figure presents the statistical distributions of key parameters for the fitting, including slope ($c_1$), intercept ($c_2$), and coefficient of determination ($R^2$), they are derived from 100,000 Monte Carlo sampling trials across three error propagation scenarios. (a), (d), (g) correspond to the slope \(c_1\) under uniform sampling, normal sampling ($\sigma$=1), and normal sampling ($\sigma$=1.96), respectively; (b), (e), (h) show the intercept \(c_2\) for the same three scenarios; (c), (f), (i) display the $R^2$ statistic. }
	\label{fig:AppendixB}
\end{figure}

In order to verify the degree of agreement between the experimental data and the relational expression given in Eq.~(\ref{Eq_Ncp}), we used logarithmic coordinates for analysis. This is because there is significant data enrichment from the low transition temperature materials, and fitting in linear coordinates would underestimate the contribution of these data points. Our fitting formula is given by
\begin{equation}
	\log_{10}(k_B T_{c}/U) = c_1 \log_{10}(1/N_{\text{CP}}) + c_2,
	\label{eq:appB}
\end{equation}
Based on the data from SI, we obtained $c_1=1.00074\pm0.00381$, $c_2=-1.70386\pm0.01435$ via fitting. The fit has a goodness of 0.997 ($R^2=0.997$) with a Residual Sum of Squares of 0.31 (RSS $=0.31$). The proximity of \(c_1\) to 1 indicates that \(k_B T_c / U\) and \(1/N_{\text{CP}}\) follow a simple linear relationship, and the proportionality coefficient $10^{c_2}\approx0.0198(3)$ corresponds to the coefficient of this linear relationship.

Since different experimental measurement methods may introduce systematic errors into the value of $U$, we systematically evaluate the impact of these uncertainties using Monte Carlo sampling. The specific methodology is as follows: (1) Using the mean values of $U$ and $N_{\text{CP}}$, we perform log-log fitting of the form relation in Eq.~(\ref{eq:appB}) to derive the fitted parameters $c_1, c_2$ and the goodness of fit $R^2$. (2) Then perform random sampling of $U$ and $N_{\text{CP}}$ around their mean values within the error range to generate databases. (3) Conduct log-log fitting on the generated data to obtain the fitted parameters $c_1, c_2$ and the goodness of fit $R^2$. (4) Repeat the sampling and data generation to obtain new fitted values of $c_1, c_2$ and $R^2$. Through multiple iterations, statistically analyze the distributions of $c_1, c_2$, and $R^2$. (5) Adjust the sampling probability distribution in step (2) (e.g., to a normal distribution), repeat the above procedures, and derive new statistical values for $c_1, c_2$, and $R^2$. (6) Quantitatively assess how uncertainties in $U$ and $N_{\text{CP}}$ affect the universal relation based on the results.

The main results of one simulation (with a sample size of 100000) are summarized in Table~\ref{tab:statistics_comparison}, where $\sigma_0$ denotes the normalized standard deviation ($\sim$Error/$\sigma_0$) of the normal distribution. A value of $\sigma_0 = 1$ signifies that the sampling range covers approximately 68\% of the experimental data, while $\sigma_0 = 1.96$ indicates that the range encompasses 95\% of the data, these are consistent with the standard confidence intervals for normal distributions. The corresponding histogram distributions of fitted parameters are presented in Fig.~\ref{fig:AppendixB}, and it can be observed that the fitted distributions of these histograms conform to the normal distribution in Fig.~\ref{fig:AppendixB}, the color lines in the figure represent the normal distribution fits. 

The results, summarized in Table~\ref{tab:statistics_comparison} demonstrate the scaling relation in Eq.~(\ref{Eq_Ncp}). The mean $R^2$ value remains high ($>0.97$) across all samplings, and the fitted exponent $c_1$ consistently deviates by less than 0.07\% from unity. Most critically, the distribution of the prefactor $\alpha=10^{c2}$ is normal, and its most probable value across the entire uncertainty space centers on 1/(16$\pi$), with deviations less than 3\%. 

In summary, it indicate that the results fit well with Eq.~(\ref{Eq_Ncp}) within the margin of statistical error.


\begin{thebibliography}{99} 
		
		\bibitem{Bednorz1986}Bednorz, J. G. \& Müller,  K. A. Possible high $T_{c}$ superconductivity in the Ba-La-Cu-O system, \textit{Z. Phys. B: Condens. Matter} \textbf{64}, 189–193 (1986).
		
		\bibitem{McMillan1968}McMillan, W. L. Transition Temperature of Strong-Coupled Superconductors, \textit{Phys. Rev.} \textbf{167}, 331 (1968)
		
		\bibitem{Cohen1972}Cohen, M. L. \& Anderson, P. W. Comments on the maximum superconducting transition temperature. \textit{AIP Conf. Proc.} \textbf{4}, 17–27 (1972).
		
		\bibitem{Bardeen1957}Bardeen, J., Cooper, L. N. \& Schrieffer, J. R. Theory of Superconductivity, \textit{Phys. Rev.} \textbf{108}, 1175 (1957).
				
		\bibitem{Keimer2015}Keimer, B., Kivelson, S., Norman, M. et al. From quantum matter to high-temperature superconductivity in copper oxides. \textit{Nature} \textbf{518}, 179–186 (2015).
		
		\bibitem{Steglich1979}Steglich, F., Aarts, J., Bredl, C. D., Lieke, W. et al. Superconductivity in the Presence of Strong Pauli Paramagnetism: CeCu$_{2}$Si$_{2}$, \textit{Phys. Rev. Lett.} \textbf{43}, 1892 (1979).
		
		\bibitem{Kamihara2008}Kamihara, Y., Watanabe, T., Hirano, M. \& Hosono, H. Iron-Based Layered Superconductor La[O$_{1-x}$F$_{x}$]FeAs (x=0.05-0.12) with T$_{c}$ = 26 K, \textit{J. Am. Chem. Soc.} \textbf{130}, 3296–3297 (2008).
		
	    \bibitem{Cao2018}Cao, Y., Fatemi, V., Fang, S. et al. Unconventional superconductivity in magic-angle graphene superlattices. \textit{Nature} \textbf{556}, 43–50 (2018). 
		
		\bibitem{Hwang2019}Li, D., Lee, K., Wang, B. Y., er al. Superconductivity in an infinite-layer nickelate, \textit{Nature} \textbf{572}, 624–627 (2019).
		
		\bibitem{Sun2023}Sun, H. et al. Signatures of superconductivity near 80 K in a nickelate under high pressure. \textit{Nature} \textbf{621}, 493-498 (2023).
		
		\bibitem{Stewart2017}Stewart G. R. Unconventional superconductivity, \textit{Adv. Phys.} \textbf{66(2)}, 75-196 (2017).
		
		\bibitem{Norman2011}Norman, M. R. The Challenge of Unconventional Superconductivity, \textit{Science} \textbf{332(6026)}, 196-200 (2011).
		
		\bibitem{Xiang2025}Xiang, T., High-temperature superconductivity: A driving force for the revolution in quantum many-body theory, \textit{Acta Physica Sinica} \textbf{74}, 077403 (2025).
		
		\bibitem{Uemura1989}Uemura, Y. J. et al. Universal correlations between T$_{c}$ and n$_{s}$/m$^{\star}$ (carrier density over effective mass) in high-T$_{c}$ cuprate superconductors. \textit{Phys. Rev. Lett.} \textbf{62}, 2317–2320 (1989).
		
		\bibitem{Homes2004}Homes, C. C. et al. A universal scaling relation in high-temperature superconductors, \textit{Nature} \textbf{430}, 539–541 (2004). 
		
		
		\bibitem{Emery1995}Emery, V., Kivelson, S. Importance of phase fluctuations in superconductors with small superfluid density. \textit{Nature} \textbf{374}, 434–437 (1995).
		
		
		\bibitem{Souma2003}Souma, S., Machida, Y., Sato, T. et al. The origin of multiple superconducting gaps in MgB$_{2}$. \textit{Nature} \textbf{423}, 65–67 (2003).
		
		\bibitem{Drozdov2025}Du, F., Drozdov, A.P., Minkov, V.S. et al. Superconducting gap of H$_{3}$S measured by tunnelling spectroscopy. \textit{Nature} (2025).
		
		\bibitem{Pena2015}Peña, O. Chevrel phases: Past, present and future, Physica C 514, 95-112 (2015).
		
		
		
		
		\bibitem{Coffman2000}Coffman, V., Kundu, J. \& Wootters, W. K., Distributed entanglement, \textit{Phys. Rev. A} \textbf{61}, 052306 (2000).
		
		\bibitem{Chen2024}Chen, Q., Wang, Z., Boyack, R., et al. When superconductivity crosses over: From BCS to BEC, Rev. Mod. Phys. 96, 025002 (2024).
		
		\bibitem{Bozovic2016}Božović, I., He, X., Wu, J. \& Bollinger, A. T., Dependence of the critical temperature in overdoped copper oxides on superfluid density, \textit{Nature} \textbf{536}, 309–311 (2016).
		
		\bibitem{Hashimoto2012}Hashimoto. K, et al. A Sharp Peak of the Zero-Temperature Penetration Depth at Optimal Composition in BaFe$_{2}$(As$_{1-x}$P$_{x}$), \textit{Science} \textbf{336}, 1554-1557 (2012). 
		
		\bibitem{Zaanen2016}Zaanen, J. Superconducting electrons go missing. \textit{Nature} \textbf{536}, 282–283 (2016).
		
		\bibitem{Banerjee2025}Banerjee, A., Hao, Z., Kreidel, M. et al. Superfluid stiffness of twisted trilayer graphene superconductors. \textit{Nature} \textbf{638}, 93–98 (2025).
		
		\bibitem{Dordevic2013}Dordevic, S., Basov, D. \& Homes, C. Do organic and other exotic superconductors fail universal scaling relations? \textit{Sci Rep} \textbf{3}, 1713 (2013).
		
		\bibitem{Kittaka2009}Kittaka, S. et al. Angular dependence of the upper critical field of Sr$_{2}$RuO$_{4}$, \textit{Phys. Rev. B} \textbf{80}, 174514 (2009).
		
		\bibitem{Mao1999}Mao, Z. Q., Mori, Y., \& Maeno, Y. Suppression of superconductivity in Sr$_{2}$RuO$_{4}$ caused by defects, \textit{Phys. Rev. B} \textbf{60}, 610 (1999).
		
		\bibitem{Steppke2017}Steppke, A., Zhao, L., Barber, M. E., et al. Strong peak in T$_c$ of Sr$_{2}$RuO$_{4}$ under uniaxial pressure. \textit{Science} \textbf{355(6321)}, eaaf9398 (2017).
		
		\bibitem{Jerzembeck2023}Jerzembeck, F., et al. Upper critical field of Sr$_{2}$RuO$_{4}$ under in-plane uniaxial pressure,\textit{ Phys. Rev. B} \textbf{107}, 064509(2023).
		
		\bibitem{Pustogow2019}Pustogow, A., et al. Constraints on the superconducting order parameter in Sr$_{2}$RuO$_{4}$ from oxygen-17 nuclear magnetic resonance, \textit{Nature} \textbf{574}, 72–75 (2019). 
		
		\bibitem{Anderson2006}Anderson, P. W., \& Ong, N. P., Theory of asymmetric tunneling in the cuprate superconductors, \textit{Phys. Chem. Solids} \textbf{67}, 1 (2006).
		
		\bibitem{Pan2000}Pan, S. H., et al. Imaging the effects of individual zinc impurity atoms on superconductivity in Bi$_{2}$Sr$_{2}$CaCu$_{2}$O$_{8+\delta}$, \textit{Nature} \textbf{403}, 746–750 (2000).
		
		\bibitem{Segawa2010}Segawa, K., Kofu, M., Lee, SH. et al. Zero-doping state and electron–hole asymmetry in an ambipolar cuprate. \textit{Nature Phys.} \textbf{6}, 579–583 (2010). 
		
		
		
		
		
		
					
		
		
		
		
		
		
		
		
		\bibitem{Nakagawa2021}Nakagawa, Y., et al. Gate-controlled BCS-BEC crossover in a two-dimensional superconductor, \textit{Science} \textbf{372(6538)}, 190-195 (2021)
						
		\bibitem{Zhou2025}Zhou, G., et al. Ambient-pressure superconductivity onset above 40 K in (La,Pr)$_3$Ni$_2$O$_7$ films. \textit{Nature} (2025). https://doi.org/10.1038/s41586-025-08755-z
		
		\bibitem{Chabaud1996}Chabaud, C., et al, Characterization of tetragonal 1:2:3 (Ca$_x$La$_{1-x}$)(Ba$_{1.75-x}$La$_{0.25+x}$)Cu$_3$O$_y$ thin films prepared by laser ablation deposition, \textit{Physica C} \textbf{261}, 33-37 (1996).
		
		
		
		
		
		
		
		
		
		
		
		\bibitem{Werthamer1966}Werthamer, N. R., Helfand, E., \& Hohenberg, P. C. Temperature and Purity Dependence of the Superconducting Critical Field, $H_{c2}$. III. Electron Spin and Spin-Orbit Effects, \textit{Phys. Rev.} \textbf{147}, 295 (1966).
		
		\bibitem{Zang2023}Zang, Q., et al. Planckian dissipation and non-Ginzburg-Landau type upper critical field in Bi2201, \textit{Sci. China Phys. Mech. Astron.} \textbf{66}, 237412 (2023).
		
		\bibitem{Wei2025}Wei, W., et al. Upper critical fields in high-$T_{c}$	superconductors, \textit{J. Phys.: Condens. Matter} \textbf{37} 143003 (2025).
		
		\bibitem{Anisimov1991}Anisimov, V. I., Zaanen J., \& Andersen O. K. Band theory and Mott insulators: Hubbard $U$ instead of Stoner $I$, \textit{Phys. Rev. B} \textbf{44}, 943 (1991).
		
		\bibitem{Dudarev1998}Dudarev, S. L., et al. Electron-energy-loss spectra and the structural stability of nickel oxide: An LSDA+U study, \textit{Phys. Rev. B} \textbf{57}, 1505 (1998).
		
		\bibitem{Antonides1977}Antonides, E., Janse, E. C., \& Sawatzky, G. A. LMM Auger spectra of Cu, Zn, Ga, and Ge. I. Transition probabilities, term splittings, and effective Coulomb interaction, \textit{Phys. Rev. B} \textbf{15}, 1669 (1977).
		
		\bibitem{Sawatzky1984}Sawatzky, G. A. \& Allen, J. W. Magnitude and Origin of the Band Gap in NiO, \textit{Phys. Rev. Lett.} \textbf{53}, 2339 (1984).
		
		\bibitem{Silva2007}Da Silva, J. L. F. et al. Hybrid functionals applied to rare-earth oxides: The example of ceria, \textit{Phys. Rev. B} \textbf{75}, 045121 (2007).
		
		\bibitem{Cococcioni2005}Cococcioni, M., \& de Gironcoli., S. Linear response approach to the calculation of the effective interaction parameters in the LDA+U method, Phys. Rev. B 71, 035105 (2005).
		
		\bibitem{Zhou2004}Zhou, F., et al. First-principles prediction of redox potentials in transition-metal compounds with LDA+U, \textit{Phys. Rev. B} \textbf{70}, 235121 (2004).
		
		\bibitem{Bach1994}Bach, V., Lieb, E. H., \& Solovej, J., P. Generalized Hartree-Fock theory and the Hubbard model, \textit{J. Stat. Phys.} \textbf{76}, 3-89 (1994).
		
		\bibitem{Aryasetiawan2004}Aryasetiawan, F., Frequency-dependent local interactions and low-energy effective models from electronic structure calculations, \textit{Phys. Rev. B} \textbf{70}, 195104 (2004).
		
		\bibitem{Vaugier2012}Vaugier, L., Jiang, H., \& Biermann, S. Hubbard $U$ and Hund exchange $J$ in transition metal oxides: Screening versus localization trends from constrained random phase approximation, \textit{Phys. Rev. B} \textbf{86}, 165105 (2012).
		
		\bibitem{Madsen2005}Madsen, G. K. H. \& Novák, P. Charge order in magnetite. An LDA+U study, \textit{EPL} \textbf{69}, 777 (2005).
		
		\bibitem{Keimer1992}Keimer, B. Magnetic excitations in pure, lightly doped, and weakly metallic La$_{2}$CuO$_{4}$, Phys. Rev. B \textbf{46}, 14034 (1992).
		
		\bibitem{Rosch2005}Rösch, O. et al. Polaronic Behavior of Undoped High-$T_{c}$ Cuprate Superconductors from Angle-Resolved Photoemission Spectra, Phys. Rev. Lett. 95, 227002 (2005).
		
		\bibitem{Shen1995}Shen, Z. -X. Photoemission Studies of High-$T_{c}$ Superconductors: The Superconducting Gap, \textit{Science} \textbf{267}, 343 (1995).
		
		\bibitem{Onsten2007}Önsten, A., et al. Probing the valence band structure of Cu$_{2}$O using high-energy angle-resolved photoelectron spectroscopy, \textit{Phys. Rev. B} \textbf{76}, 115127 (2007).
		
		\bibitem{Shen1987}Shen, Z. -X., et al. Anderson Hamiltonian description of the experimental electronic structure and magnetic interactions of copper oxide superconductors, \textit{Phys. Rev. B} \textbf{36}, 8414 (1987). 
		
		\bibitem{Eskes1990}Eskes, H., Tjeng, L. H., \& Sawatzky, G. A., Cluster-model calculation of the electronic structure of CuO: A model material for the high-$T_{c}$ superconductors, \textit{Phys. Rev. B} \textbf{41}, 288 (1990).
		
		\bibitem{Hybertsen1989}Hybertsen, M. S., Schlüter, M., \& Christensen, N. E., Calculation of Coulomb-interaction parameters for La$_{2}$CuO$_{4}$ using a constrained-density-functional approach, \textit{Phys. Rev. B} \textbf{39}, 9028 (1989).
		
		\bibitem{Fujimori1987}Fujimori, A., et al. Spectroscopic evidence for strongly correlated electronic states in La-Sr-Cu and Y-Ba-Cu oxides, Phys. Rev. B 35, 8814(R) (1987).
		
		\bibitem{Zaanen1986}Zaanen, J., Westra, C., \& Sawatzky, G. A. Determination of the electronic structure of transition-metal compounds: 2p x-ray photoemission spectroscopy of the nickel dihalides, \textit{Phys. Rev. B} \textbf{33}, 8060 (1986).
		
		\bibitem{Marel1988}van der Marel, D., van Elp, J., Sawatzky, G. A., \& Heitmann, D. X-ray photoemission, bremsstrahlung isochromat, Auger-electron, and optical spectroscopy studies of Y-Ba-Cu-O thin films, \textit{Phys. Rev. B} \textbf{37}, 5136 (1988).
		
		\bibitem{Fuggle1988}Fuggle, J. C., et al. Valence bands and electron correlation in the high-$T_{c}$ superconductors, \textit{Phys. Rev. B} \textbf{37}, 123 (1988).
		
		\bibitem{Sheshadri2023}Sheshadri, K., Malterre, D., Fujimori, A., \& Chainani, A. Connecting the one-band and three-band Hubbard models of cuprates via spectroscopy and scattering experiments, Phys. Rev. B 107, 085125 (2023).
		
		\bibitem{Moore2024}Moore, G. C., et al. High-throughput determination of Hubbard $U$ and Hund $J$ values for transition metal oxides via the linear response formalism, Phys. Rev. Materials 8, 014409 (2024).
		
		\bibitem{Anisimov1997}Anisimov, V. I., Aryasetiawan, F., \& Lichtenstein, A. I. First-principles calculations of the electronic structure and spectra of strongly correlated systems: the LDA+ U method, \textit{J. Phys.: Condens. Matter} \textbf{9}, 767 (1997).
		
		\bibitem{Schrieffer1966}Schrieffer, J. R., \& Wolff, P. A. Relation between the Anderson and Kondo Hamiltonians, \textit{Phys. Rev.} \textbf{149}, 491 (1966).
		
		\bibitem{Anderson1961}Anderson, P. W. Localized Magnetic States in Metals, \textit{Phys. Rev.} \textbf{124}, 41 (1961).
		
	
		

		
		
		
	\end{thebibliography}

\end{document}